\documentclass[conference]{IEEEtran}
\IEEEoverridecommandlockouts
\usepackage{cite}
\usepackage{amsmath,amssymb,amsfonts}
\usepackage{graphicx}
\usepackage{textcomp}
\usepackage{xcolor}
\usepackage{amsmath, amssymb}
\usepackage{algorithm}
\usepackage{algpseudocode}
\usepackage{physics}
\usepackage{float} 
\usepackage{subcaption}
\usepackage{booktabs}
\usepackage{multirow}
\usepackage{wrapfig} 
\usepackage{array}    
\usepackage{booktabs} 
\usepackage{caption}  
\usepackage{adjustbox}

\def\BibTeX{{\rm B\kern-.05em{\sc i\kern-.025em b}\kern-.08em
    T\kern-.1667em\lower.7ex\hbox{E}\kern-.125emX}}
\begin{document}

\title{An Interpretable Federated Learning Control Framework Design
for Smart Grid Resilience}

\author{
  \IEEEauthorblockN{
    Ibrahim Shahbaz\IEEEauthorrefmark{1},~\IEEEmembership{Student Member,~IEEE},
    Eman Hammad\IEEEauthorrefmark{1},~\IEEEmembership{Senior Member,~IEEE}, and\\
    Abdallah Farraj\IEEEauthorrefmark{2},~\IEEEmembership{Senior Member,~IEEE}
  }
\IEEEauthorblockA{\IEEEauthorrefmark{1}are with iSTAR Lab at Texas A\&M University, College Station, TX 77840, USA.}
\IEEEauthorblockA{\IEEEauthorrefmark{2}is with Texas A\&M University Texarkana.}
}

\maketitle

\begin{abstract}

Power systems remain highly vulnerable to disturbances and cyber-attacks, underscoring the need for resilient and adaptive control strategies. In this work, we investigate a data-driven Federated Learning Control (FLC) framework for transient stability resilience under cyber-physical disturbances. The FLC employs interpretable neural controllers based on the Chebyshev Kolmogorov-Arnold Network (ChebyKAN), trained on a shared centralized control policy and deployed for distributed execution. Simulation results on the IEEE 39-bus New England system show that the proposed FLC consistently achieves faster stabilization than distributed baselines at moderate control levels (10\%--60\%), highlighting its potential as a scalable, resilient, and interpretable learning-based control solution for modern power grids.

\end{abstract}


\section{Introduction}

Smart grids (SGs), modeled as cyber-physical systems (CPS), integrate sensing, communication, and control with traditional power networks to enable reliable electricity delivery~\cite{ruiz2014integration}. Yet they remain exposed to cyber threats and disturbances that compromise stability and reliability~\cite{PG_reselience_adverseries,impact_cyber_attacks}. This has driven resilient-by-design strategies to anticipate, absorb, adapt, and recover from disruptions. Intelligent control schemes that couple physical and cyber dynamics are essential to such strategies~\cite{lun2019state}. Advances in artificial intelligence (AI), especially machine learning (ML) and deep learning (DL), offer new opportunities to enhance resilience in SGs~\cite{review_ML_PS_reselience}. This work applies advanced AI methods to transient stability control in transmission networks.


Traditional SG control schemes—centralized, distributed, and hybrid—improve resilience through sensing and coordination but involve trade-offs among resilience, performance, and scalability~\cite{nandanoori2024empowering}. Centralized schemes achieve high performance using system-wide PMU data but create a vulnerable single point of failure. Distributed schemes avoid this dependence yet lack global visibility, limiting effectiveness under disturbances. Hybrid approaches attempt to balance these drawbacks by switching between centralized and distributed modes based on specific criteria(e.g. latency thresholds)~\cite{re_feedback_linearization,farraj2016cyber}.

AI and reinforcement learning (RL) have emerged as promising data-driven approaches to enhance control and resilience in SGs~\cite{review_ML_PS_reselience}. These methods enable model-free, intelligent decision-making that enhances stability, optimizes resource allocation, and supports self-healing in both microgrids (MGs) and large-scale systems~\cite{ AI_MG_Resilience}. RL is particularly suited for managing non-linear dynamics and uncertainties in MG control~\cite{ RL_MG_Control}, while AI techniques support predictive maintenance, anomaly detection, and real-time optimization~\cite{ AI_Hierarchical_control_MGs}. However, practical deployment faces challenges, including reliance on high-quality data, limited interpretability and generalizability of learning models, and difficulties in adapting to uncertain and dynamic environments.


Limitations of traditional control and data-driven methods motivate exploring innovative learning-based techniques to enhance SG cyber-physical security and resilience. Federated learning (FL) has emerged as a promising framework for distributed intelligence, privacy, and adaptability in control systems~\cite{weber2024combining}. By enabling collaborative training while keeping data local and sharing only model parameters, FL preserves confidentiality, reduces communication overhead, and mitigates data dependency, making it well-suited for networked CPS. Integrating FL into control loops further improves adaptability to dynamic conditions and model generalization~\cite{advancing_PS_FL}. Scientific machine learning (SciML) has likewise gained attention as an explainable paradigm that combines data-driven models with physical principles to improve interpretability and trust. Within this context, Kolmogorov–Arnold Networks (KANs)~\cite{KANs} offer a novel neural architecture with greater expressivity and interpretability than conventional deep neural networks.

A comprehensive review of FL applications in power systems is provided in~\cite{advancing_PS_FL}, spanning generation, distribution , microgrids (MGs) , and cybersecurity. FedGrid~\cite{gupta2023fedgrid} introduces a secure FL framework for privacy-preserving forecasting of renewable generation and load. In~\cite{FRL_decentrelized}, a federated RL (FRL) scheme enables decentralized voltage control in distribution networks without sharing raw data, improving scalability and privacy. A vertical FRL strategy using a federated soft actor-critic (FedSAC) algorithm is proposed in~\cite{mukherjee2024resilient} for resilient MG control under adversarial attacks. Validated via co-simulationand hardware-in-the-loop testing, the vertical FRL strategy achieves robust voltage stabilization and effective defense against cyber threats.

\begin{figure}[!ht]
\centering
\includegraphics[width=0.95\columnwidth]{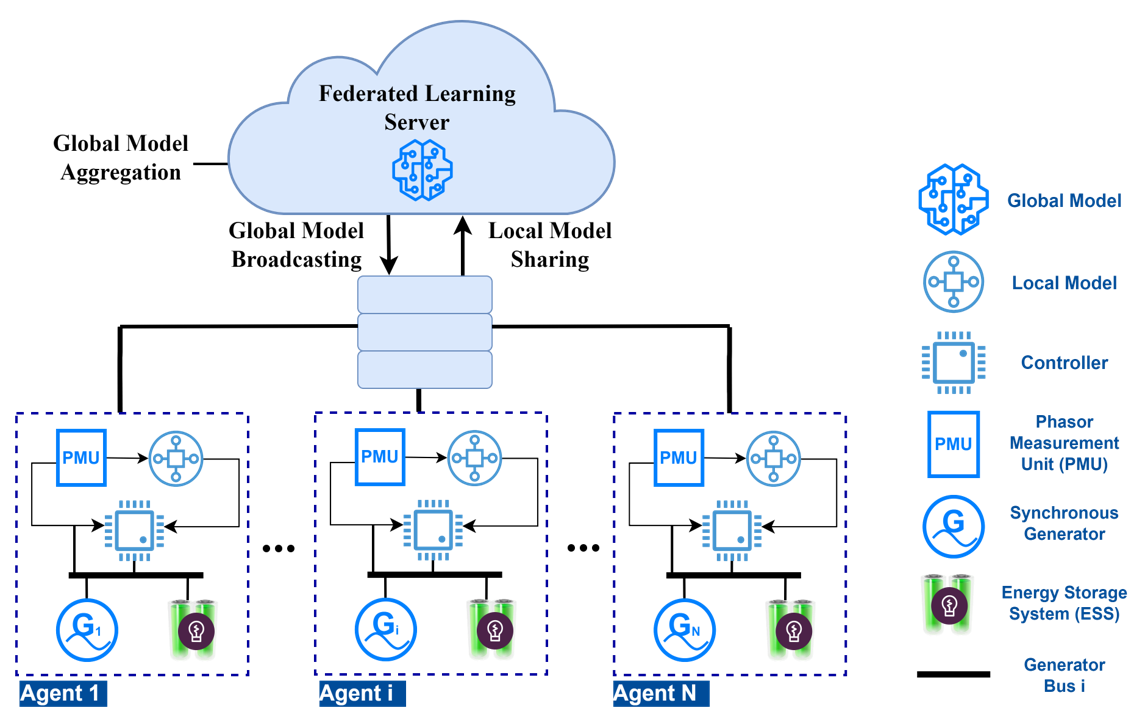}
\caption{Proposed federated learning control architecture.}
\label{fig:FL_SG}
\end{figure}

In this work, we propose a FL-based control (FLC) framework for transient stability in SG transmission networks. The framework integrates a distributed, interpretable neural controller—Chebyshev KAN (Cheby KAN)~\cite{chebyKAN}—that is continuously trained on centralized control actions in a federated manner. By approximating a global centralized policy through collaborative ML, each local controller benefits from system-wide experience without sharing raw PMU data, thereby preserving privacy and reducing communication overhead. The main contributions in this work are listed as follows:
\begin{itemize} 
    \item  Development of a transient stability neural controller for SGs that combines FL with hybrid centralized--distributed control schemes.
    \item Improved interpretability of the neural controller decision-making process by employing Cheby KANs, which utilize edge-wise learnable polynomial functions. 
    \item Implementation and evaluation of the framework on the IEEE 39-bus New England system, demonstrating generalization to unseen fault scenarios.  
\end{itemize}

\vspace{-0.1in}
\section{Background - Chebyshev Kolmogorov-Arnold Networks}\label{Cheby_KANs}
\vspace{-0.01in}
KANs are a recently proposed class of neural architectures inspired by the Kolmogorov–Arnold representation theorem, which states that any multivariate continuous function can be represented as compositions of univariate continuous functions and additions~\cite{KANs}. Unlike multilayer perceptrons (MLPs), which place parameters on nodes with fixed activations, KANs assign learnable functions to edges, typically parameterized by basis expansions such as B-splines. This edge-centric design allows flexible approximation of non-linear transformations, enabling KANs to capture high-dimensional functions with fewer parameters and greater interpretability. Cheby KANs extend this idea by employing Chebyshev polynomials instead of B-splines, offering improved numerical stability and stronger non-linear approximation while preserving the interpretability of edge-parameterized functions~\cite{chebyKAN}.



\section{Problem Setup}\label{Problem_setup}
\label{sec:setup}
The general setup adopted in this work for transient stability control of SG transmission network can be modeled as a multi-agent system, as illustrated in Fig.~\ref{fig:FL_SG}, where each agent is composed of the following components:  1) A synchronous generator; 2) a PMU device providing local measurements, 3) an energy storage system (ESS) capable of fast active power charging/discharging, and 4) a local intelligent controller agent located at each generator bus.

The system dynamics of the interconnected synchronous generators are governed by the swing equation~\cite{PS_Analysis}. It describes the time evolution of a generator's rotor angle \( \delta \) and angular frequency \( \omega \), which are influenced by mechanical and electrical powers. This equation is essential for studying the transient stability of the system, especially under fault conditions or system disturbances. In a typical power system, the dynamics of generator \(i\) \(\forall i \in \{1, \dots, N\}\) are governed by \(\dot{\delta_i} = \omega_i\), and 
\begin{equation}\label{eqn: swing 2}
M_i \dot{\omega_i} = -D_i \omega_i + (P_{m,i} - P_{e,i}),
\end{equation}
where $\delta_i$ is the rotor angle of the generator, $\omega_i$ is the normalized angular frequency of the generator (in radians per second), \( M_i \) is the inertia of the generator, \( D_i \) is the damping coefficient, \( P_{m,i} \) is the mechanical power input to the generator, \( P_{e,i} \) is the electrical power output of generator \( i \), and \( \dot{\delta}_i \) and \( \dot{\omega}_i \) are the derivatives of \( \delta_i \) and \( \omega_i \) with respect to time. The electrical power of Generator \( i \) is defined as:
\begin{equation*}\label{eqn: electrical power}
P_{e,i} = \sum_{k=1}^{N} |E_i|\,|E_k| \left[ G_{ik} \cos(\delta_i - \delta_k) + B_{ik} \sin(\delta_i - \delta_k) \right], 
\end{equation*}
where \(|E_i|\,|E_k|\) are the internal voltages of generators \(i,k\) respectively, and \( G_{ik} = G_{ki} \geq 0 \) and \( B_{ik} = B_{ki} > 0 \) are the reduced equivalents of conductance and susceptance from a kron-reduction based method~\cite{kron_reduction} that represents the power grid as system of interconnected generators.


We assume that the default control scheme stabilizing the power system after disturbances is the centralized parametric feedback linearization (CPFL) controller scheme~\cite{re_feedback_linearization}. In CPFL, the control action is computed as the negation of the difference between the accelerating power term \(P_{a,i}\) and the decentralized PFL controller (DPFL) action term, \(P_{d,i}\), as follows: \begin{equation}\label{eqn: CPFL Pu}
P_{u,i} = -\left( P_{a,i} - P_{d,i}  \right),
\end{equation}
 where $P_{a,i} = P_{m,i} - P_{e,i}$. The DPFL controller action only utilizes the local PMU measurements and is computed as follows:  \begin{equation}\label{eqn: DPFL Pu}
 P_{d,i} = -\left( \alpha_i \omega_i + \beta_i (\delta_i - \delta_i^*) \right),
 \end{equation}
 where 
$\alpha_i,\beta_i \geq 0$ are the frequency and phase stability parameters respectively, and $\delta_i^*$ is the desired generator phase. The amount of power injection/absorption at each generator \(i\) bus is formulated to cancel all non-linear terms in~\eqref{eqn: swing 2}. The value of \(P_{u,i} \)  is a scalar commanding the charging/discharging amount of the local ESS, where a negative value indicates absorbing active power from the generator bus by charging the ESS, while a positive value indicates injecting active power by discharging the ESS. 

\section{Federated Learning control framework}\label{FLC_framework}
\label{sec:FLC}

The FLC multi-agent framework, shown in Fig.~\ref{fig:FL_SG}, consists of \(N\) intelligent agents and a FL server. Each agent hosts a local neural controller, acting as an edge device co-located with a fast-acting ESS at a generator bus. Local controllers periodically send their learned parameters, \(\theta\), representing local control policies, to the FL server. The server aggregates these parameters using an FL strategy to form a global control policy that reflects the collective generator dynamics, and then broadcasts the updated policy back to all local controllers.

Each local neural controller is a Cheby KAN model, \(f_{\theta_i}(\omega_i, \Delta\delta_i, t)\), that is configured to approximate the non-linear \(P_{a,i}\) term in the centralized control action formulated in~\eqref{eqn: CPFL Pu}, given local PMU measurements only (i.e. \(\omega_i\),  \(\Delta\delta_i =\delta_i - \delta_i^* \)). Hence, for each time step \(t\), by optimizing the local Cheby KAN parameters \(\theta_{i}\) to minimize 
loss function \(\mathcal{L}_{i}(\theta_{i})=\left\| f_{\theta_i}(\omega_i, \delta_i, t) - P_{a,i} \right\|^2\), where each local neural controller is continuously trained to approximate the non-linear \(P_{a,i}\) term.

The FL server learns the global control policy, \(\theta_{G}\), by aggregating local control policies shared by each local agent in the form of network parameters \(\theta_{i}\). 
The aggregation is performed at an aggregator node, which deploys the FedAvg~\cite{FedAvg} algorithm to produce an improved global model at each iteration as:
\vspace{-0.1in}
\begin{equation}\label{eqn: FedAVG}
    \theta_{G} = \frac{1}{N} \sum_{i=1}^N \theta_i.
\end{equation}

The updated global policy is broadcast to all local neural controllers for the next learning round with globally informed parameters. Through FLC, this policy is continuously refined and shared, operating persistently until intervention is required to complement or replace the centralized scheme. Each local controller, paired with a synchronous generator, supports system recovery using only local PMU data. To approximate near-optimal centralized control, each controller adjusts its distributed action with the addition of the approximated \(\hat{P}_{a,i}\). Hence, the approximated control action for each controller is given as:
\vspace{-0.5em}
\begin{equation}\label{eqn: FLC Pu}
\hat{P_{u,i}} = -\left( \hat{P}_{a,i} - P_{d,i}  \right).
\end{equation}
\vspace{-0.1em}FLC resembles a “learning-based feedback linearization”, where non-linear terms in~\eqref{eqn: swing 2} are canceled by learning from centralized actions, without requiring wide-area PMU data in real-time. The overall logic of the proposed FLC framework is outlined in Algorithm~\ref{alg:FLC}.

\begin{algorithm}
\scriptsize{
\caption{FLC of neural controllers for centralized action emulation}
\label{alg:FLC}
\begin{algorithmic}[1]
\State Initialize local neural controller parameters $\theta_i$ \(\forall i \in \{1, \dots, N\}\)
\While{True}
    \For{agent $G_{i}$ \(\forall i \in \{1, \dots, N\}\) in parallel}
        \State Receive current local PMU data $\omega_i$, $\Delta\delta_i$, at time $t$
        \State Query centralized controller to obtain $P_{a,i}$
        \State Update local parameters $\theta_i$ using gradient descent
    \EndFor
    \State Aggregate models at central server using FedAvg~\eqref{eqn: FedAVG}
    \State Broadcast updated global model $\theta_{G}$ to all agents
    \For{agent $G_{i}$ \(\forall i \in \{1, \dots, N\}\)}
        \If{generator node $i$ is down}
            \State deploy local neural controller: 
            \[
            \hat{P_{a,i}} \gets f_{\theta_j}(\omega_j, \delta_j, t)
            \]
            \[
            \hat{P_{u,i}}  \gets eq ~\eqref{eqn: FLC Pu}
            \]
        \EndIf
    \EndFor
\EndWhile
\end{algorithmic}
}
\end{algorithm}

\section{Numerical Results}\label{numerical_Results}

\subsection{Experimental Setup}

\begin{figure}[!t]
    \centering
    \includegraphics[page=1, width=0.8\columnwidth, trim=0mm 1mm 0mm 0.5mm, clip]{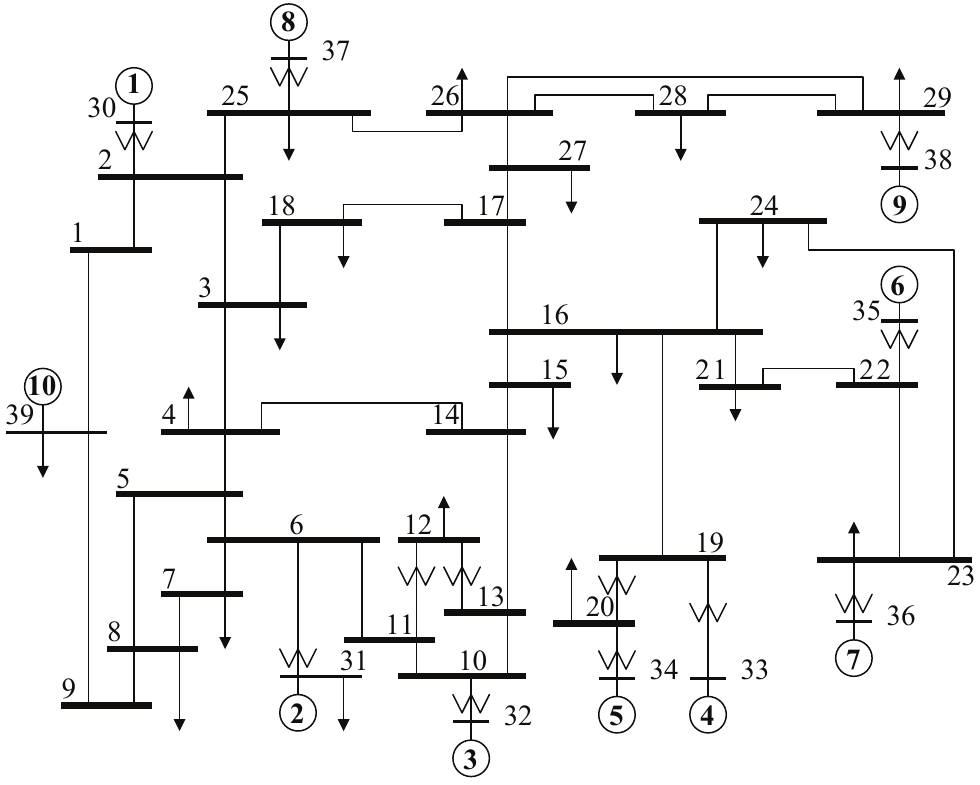}
    \caption{IEEE-39 New England bus system single-line diagram.}
    \label{fig:ieee39_sld}
\end{figure}

The FLC framework is implemented using a client--server architecture via the Flower FL platform~\cite{flowerFL}. Each agent node runs a PyTorch-based Cheby KAN model built with \texttt{Deep-KAN} (v0.0.4). After hyperparameter tuning, the selected architecture includes two layers: the first maps three input features to 32 hidden nodes using degree-five Chebyshev polynomials; the second outputs a scalar via another degree-five expansion. Models are trained on MATLAB-generated transient stability simulation data from the IEEE 39-bus, 10-generator system (Fig.~\ref{fig:ieee39_sld}) and integrated into the same simulation for testing. Experiments used a Debian GNU/Linux VM (kernel 6.1.0-33-amd64) with 8 cores @2\,GHz, 64\,GB RAM, PyTorch v2.6.0+cpu, and MATLAB R2024b.

\begin{wraptable}[8]{r}{0.25\textwidth} 
\vspace{-2pt} 
\centering
\small{
\setlength{\tabcolsep}{1pt}    
\renewcommand{\arraystretch}{0.8} 
\caption{Fault Details}
\label{tab:fault_details}
\begin{tabular}{@{}c c c@{}}  
\toprule
\textbf{Fault} & \textbf{Faulted Bus} & \textbf{Tripped Line} \\
\midrule
F1 & 17 & 17--18 \\
F2 & 11 & 10--11 \\
F3 & 22 & 21--22 \\
F4 & 29 & 28--29 \\
F5 & 5  & 5--8   \\
\bottomrule
\end{tabular}
\vspace{6pt}
}
\end{wraptable}

To train and evaluate the FLC framework, a dataset was generated for 10-generator dynamics (\(\omega, \Delta\delta, t\)) and centralized control action (\({P_{a}}\)) under a three-phase fault (F1), initiated at $t_{\text{fault}} = 0.5$~s and cleared at $t_{\text{clearing}} = 0.75$~s. Data were sampled at $\Delta t = 10^{-3}$~s over $t_{\text{max}} = 100$~s. Federated training ran for 20 rounds (398.89 seconds total), yielding an average mean squared error (MSE) of 4.26. The FLC framework is evaluated for resilience and generalization under various three-phase short-circuit faults, as detailed in Table~\ref{tab:fault_details}.

In all experiments, $\alpha_i = 0.5$ and $\beta_i = 0.005$ are fixed. Stability time is defined as the duration to regulate $\omega_i$ within $|\omega_i| \leq 0.01$. Governor control is disabled to assess the FLC’s standalone stabilization capability under the specified disruptions.

\subsection{Generalizability Across Faults} \label{case_study}


This study evaluates the generalization of the proposed FLC scheme trained solely on fault F1 and tested on unseen faults (F2--F5). Distributed control penetration levels range from \(10\%\) to \(80\%\), where a subset of generators is governed by FLC and the remainder by CPFL. For instance, FLC~80\% maps to \(G_1\)--\(G_8\) controlled under FLC and \(G_9\)--\(G_{10}\) under CPFL~20\%. This penetration definition is consistently applied to assess the generalizability, adaptability, and performance of the FLC framework relative to DPFL benchmarks. The results of this experiment are reported in Table~\ref{tab:flc_dpfl_comparison}, which presents distributed FLC vs. DPFL benchmarking, while Table~\ref{tab:cpfl_only} reports the complementary CPFL performance.


Table~\ref{tab:flc_dpfl_comparison} reports the average stability times of FLC and DPFL across penetration levels (10\%--80\%), where each value represents the average stability time of its controlled generators (e.g., FLC~20\% averages \(G_{1}\) and \(G_{2}\)). Despite being trained only on fault F1, the FLC generalizes reasonably well across unseen faults. At lower penetrations (10\%--50\%), FLC outperforms DPFL in most scenarios (F2, F3, F5), while DPFL shows advantage primarily in F4. However, beyond 60\% penetration, FLC performance declines, with DPFL becoming superior past 70\%.

\begin{figure*}[!t]
    \centering

    \begin{subfigure}{0.32\textwidth}
        \centering
        \includegraphics[width=\linewidth]{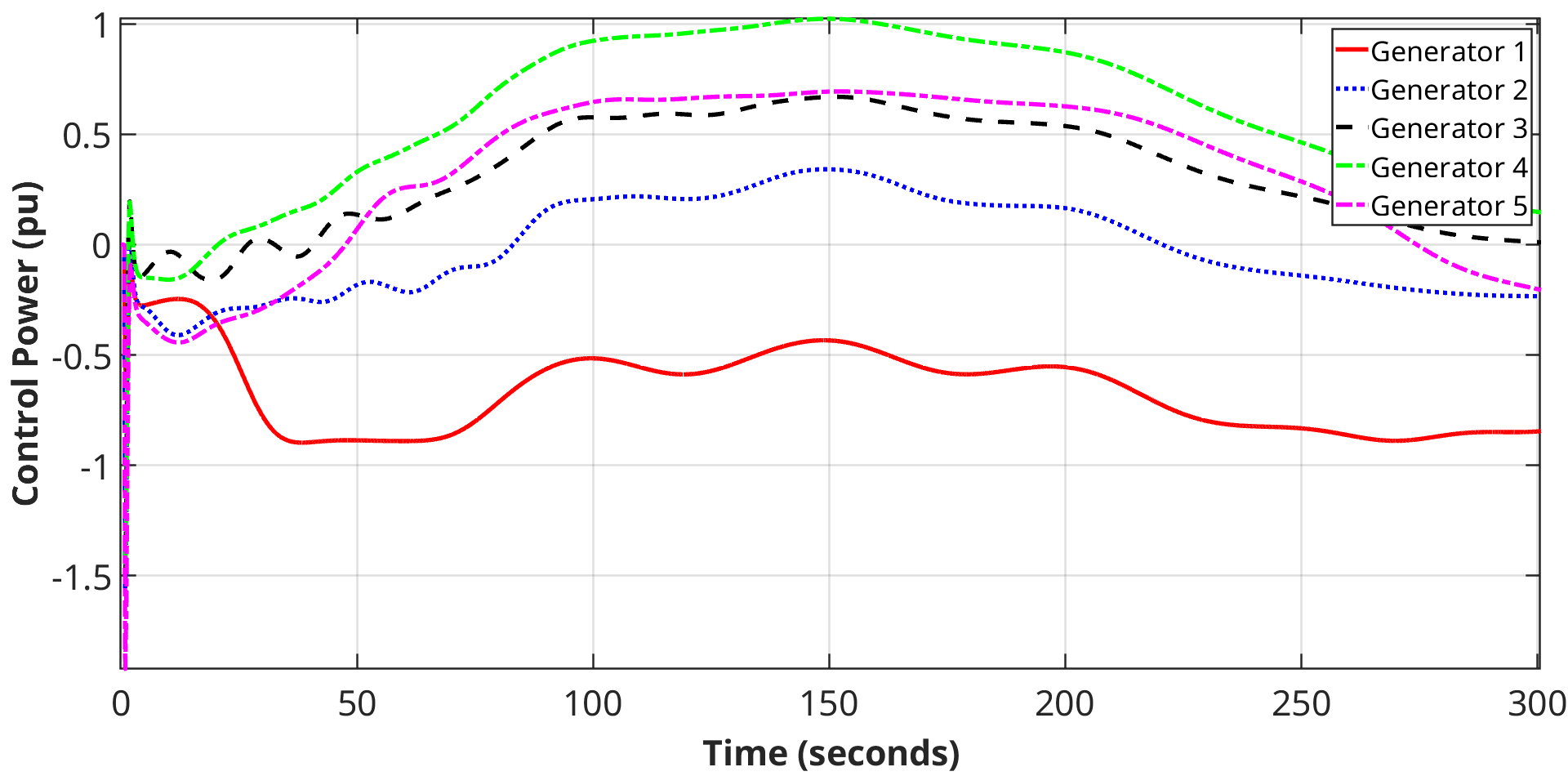}
        \caption{FLC: control power}
        \label{fig:flc_f2_pu}
    \end{subfigure}
    \hfill
    \begin{subfigure}{0.32\textwidth}
        \centering
        \includegraphics[width=\linewidth]{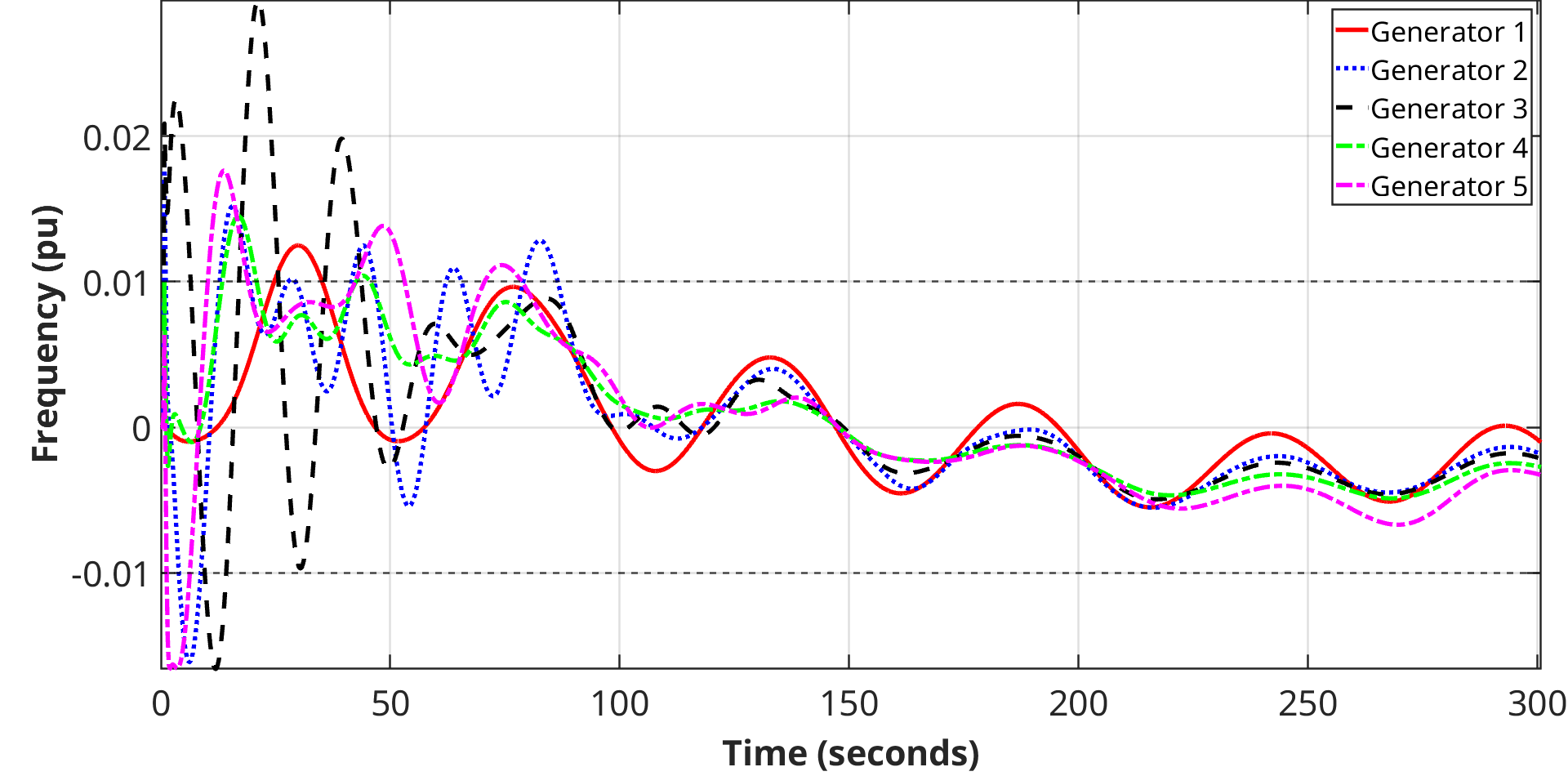}
        \caption{FLC: angular frequency ($\omega$)}
        \label{fig:flc_f2_omega}
    \end{subfigure}
    \hfill
    \begin{subfigure}{0.32\textwidth}
        \centering
        \includegraphics[width=\linewidth]{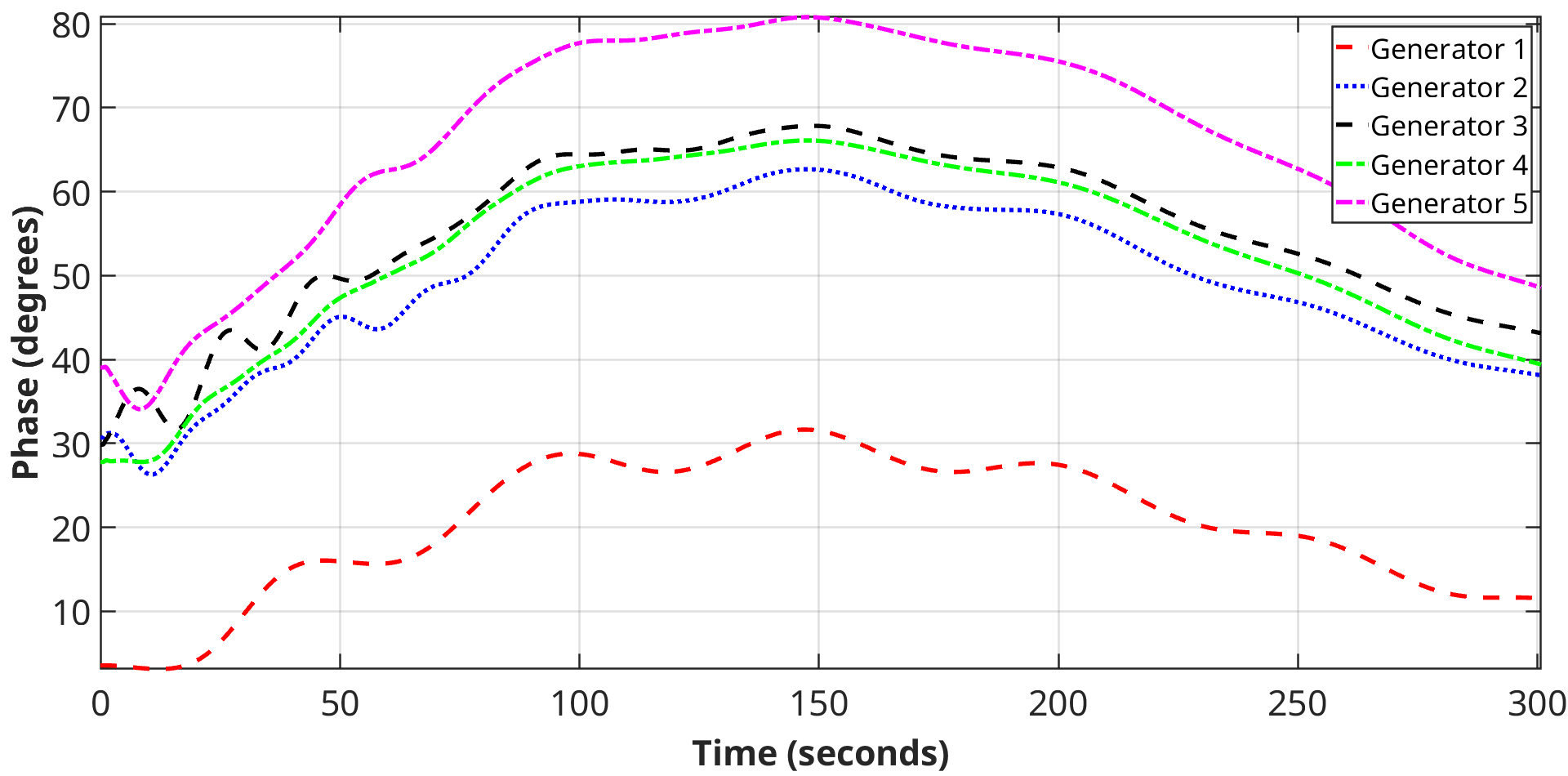}
        \caption{FLC: rotor angle (\(\delta\))}
        \label{fig:flc_f2_phase}
    \end{subfigure}

    \vspace{0.6em}

    \begin{subfigure}{0.32\textwidth}
        \centering
        \includegraphics[width=\linewidth]{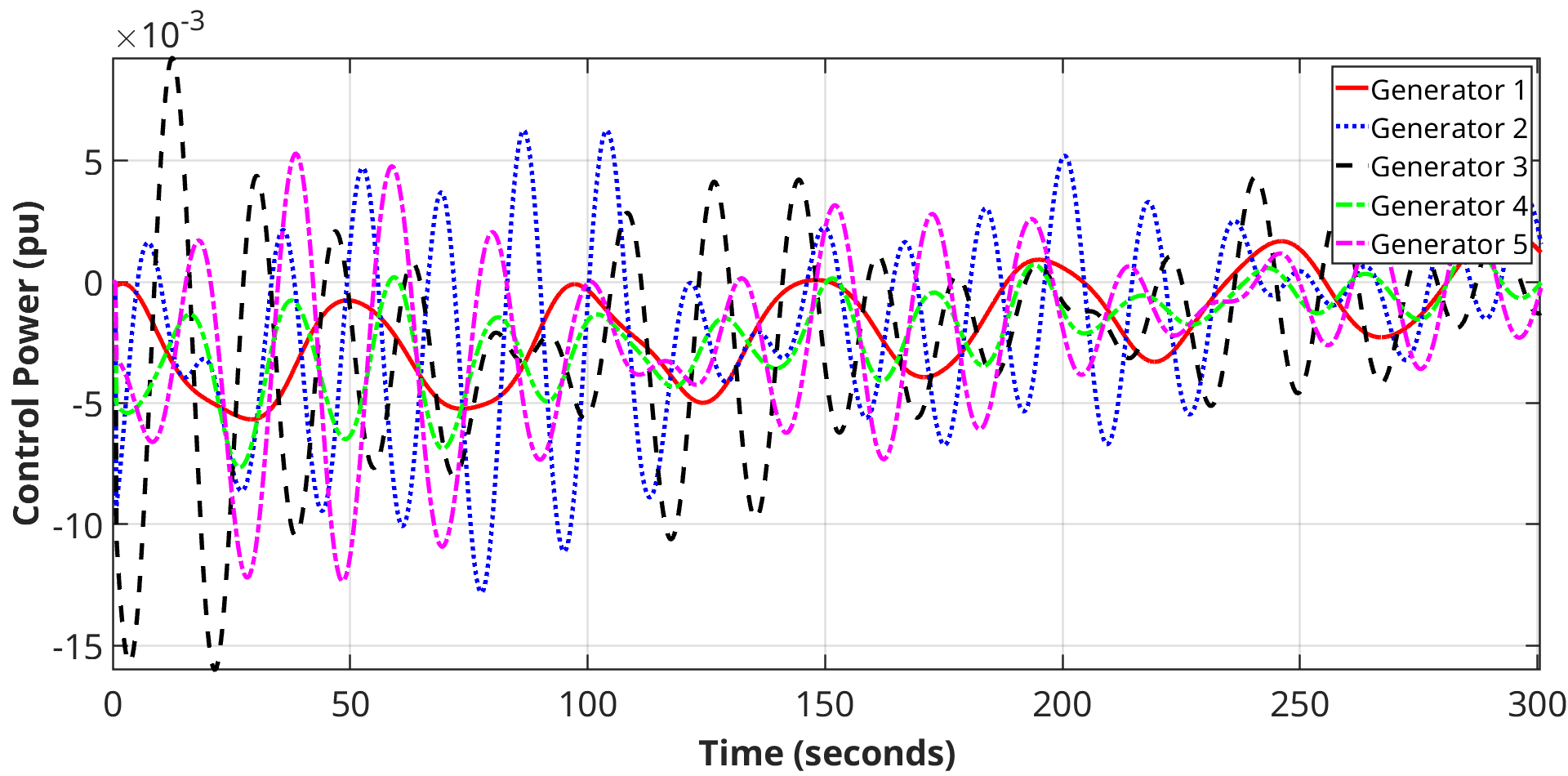}
        \caption{DPFL: control power}
        \label{fig:dpfl_f2_pu}
    \end{subfigure}
    \hfill
    \begin{subfigure}{0.32\textwidth}
        \centering
        \includegraphics[width=\linewidth]{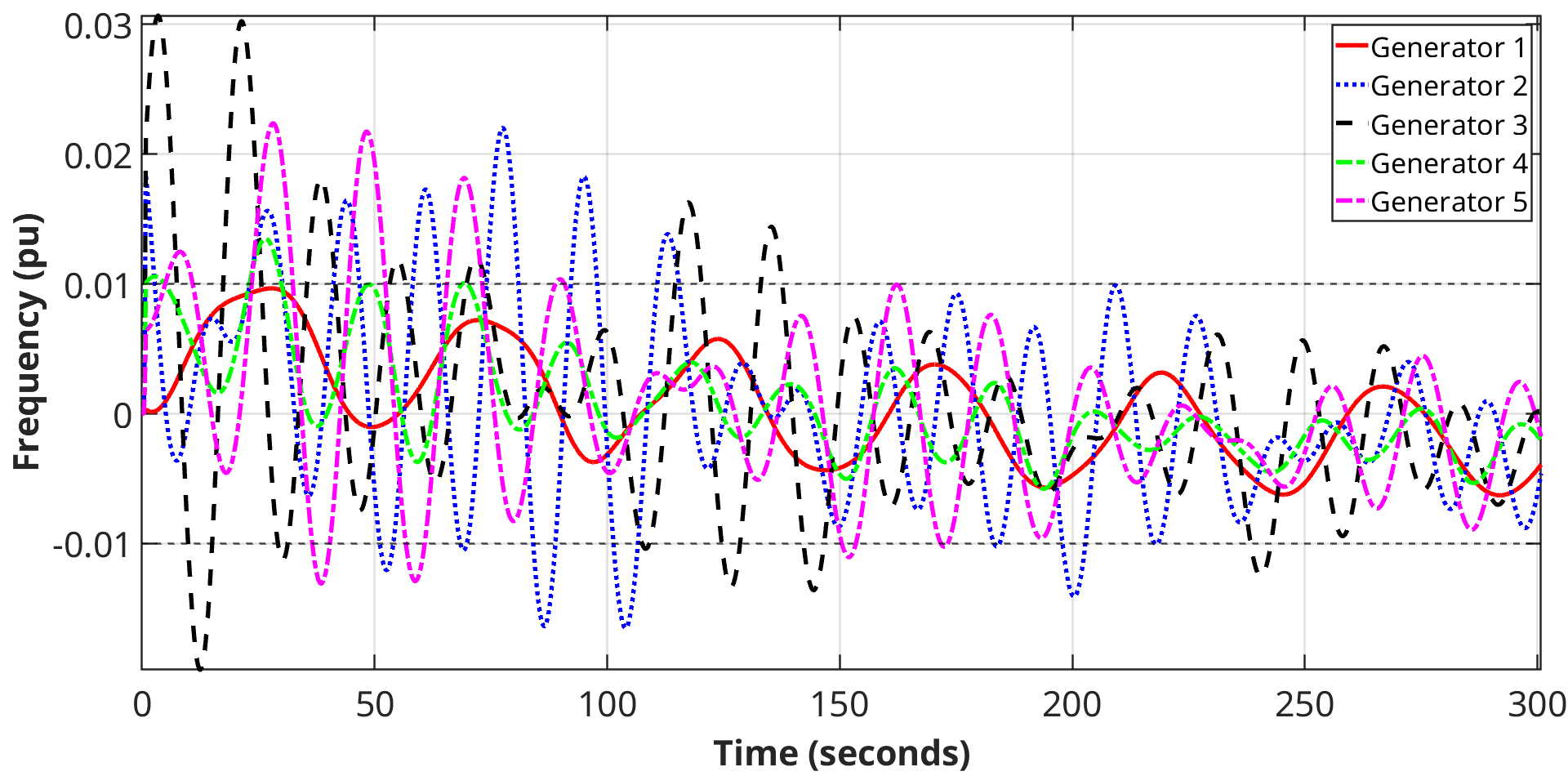}
        \caption{DPFL: angular frequency ($\omega$)}
        \label{fig:dpfl_f2_omega}
    \end{subfigure}
    \hfill
    \begin{subfigure}{0.32\textwidth}
        \centering
        \includegraphics[width=\linewidth]{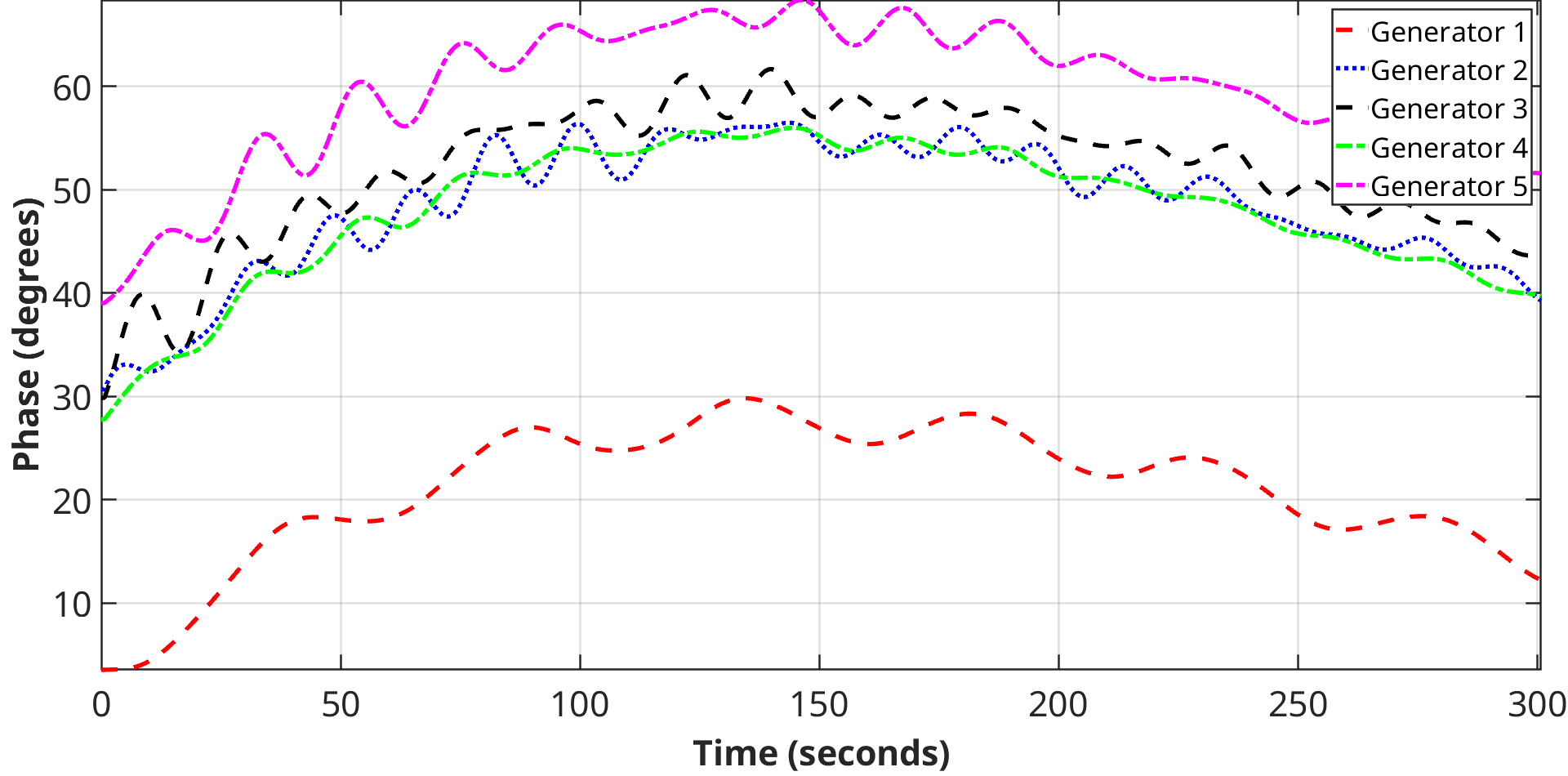}
        \caption{DPFL: rotor angle (\(\delta\))}
        \label{fig:dpfl_f2_phase}
    \end{subfigure}

    \caption{Dynamic responses under fault F2 controlled via FLC 50\% and DPFL 50\%.}
    \label{fig:flc_vs_dpfl_f2}
\end{figure*}

The responses for FLC~50\% and DPFL~50\% under fault F2 are illustrated in Fig.~\ref{fig:flc_vs_dpfl_f2}. FLC~50\% quickly restores all controlled generators to the angular frequency margin $|\omega_i| \leq 0.01$, achieving stability significantly faster than DPFL~50\%, which exhibits oscillations and delayed synchronization. Additionally, FLC~50\% provides a smoother power control profile, avoiding the high-frequency fluctuations of DPFL~50\% which can accelerate battery degradation.

\begin{table}[!t]
\centering
\scriptsize
\caption{Comparison of average stability times (seconds) between FLC and DPFL across different distributed control penetration levels under various fault scenarios.}
\label{tab:flc_dpfl_comparison}
\adjustbox{max width=\columnwidth}{
\begin{tabular}{c|cc|cc|cc|cc}
\toprule
\multirow{2}{*}{\textbf{Distributed Control Level}} & \multicolumn{2}{c|}{\textbf{F2}} & \multicolumn{2}{c|}{\textbf{F3}} & \multicolumn{2}{c|}{\textbf{F4}} & \multicolumn{2}{c}{\textbf{F5}} \\
& \textbf{FLC} & \textbf{DPFL} & \textbf{FLC} & \textbf{DPFL} & \textbf{FLC} & \textbf{DPFL} & \textbf{FLC} & \textbf{DPFL} \\
\midrule
10\% & \textbf{74.3} & 413.2 & \textbf{28.2} & 71.2 & \textbf{0.0} & \textbf{0.0} & \textbf{76.1} & 460.2 \\
20\% & \textbf{78.2} & 239.4 & 72.1 & \textbf{53.1} & 15.9 & \textbf{0.0} & \textbf{75.2} & 82.9 \\
30\% & \textbf{45.5} & 201.3 & 84.5 & \textbf{77.5} & 34.3 & \textbf{8.6} & \textbf{38.1} & 71.3 \\
40\% & \textbf{44.2} & 149.3 & \textbf{81.1} & 136.3 & 29.8 & \textbf{7.0} & \textbf{38.6} & 99.5 \\
50\% & \textbf{57.2} & 140.2 & \textbf{117.8} & 190.8 & 75.7 & \textbf{9.6} & \textbf{75.3} & 105.3 \\
60\% & \textbf{64.0} & 128.5 & \textbf{124.5} & 272.3 & 151.2 & \textbf{9.0} & 78.5 & \textbf{69.4} \\
70\% & 369.4 & \textbf{115.1} & 400.1 & \textbf{383.3} & 389.3 & \textbf{58.4} & 379.8 & \textbf{77.4} \\
80\% & 499.6 & \textbf{100.2} & 1000.1 & \textbf{511.3} & 399.8 & \textbf{63.2} & 399.8 & \textbf{59.7} \\
\bottomrule
\end{tabular}}
\end{table}

\begin{table}[!t]
\centering
\scriptsize
\caption{Average stability times (seconds) for CPFL under various fault scenarios across different centralized control penetration levels.}
\label{tab:cpfl_only}
\begin{tabular}{c|c|c|c|c}
\toprule
\textbf{Centralized Control Level} & \textbf{F2} & \textbf{F3} & \textbf{F4} & \textbf{F5} \\
\midrule
90\% & 12.8 & 16.6 & 9.6 & 13.9 \\
80\% & 8.1 & 18.7 & 10.8 & 7.8 \\
70\% & 0.0 & 21.4 & 12.3 & 0.6 \\
60\% & 0.0 & 21.6 & 14.4 & 0.0 \\
50\% & 0.0 & 25.9 & 17.2 & 0.0 \\
40\% & 0.0 & 14.4 & 21.5 & 0.0 \\
30\% & 0.0 & 0.0 & 28.6 & 0.0 \\
20\% & 0.0 & 0.0 & 43.2 & 0.0 \\
\bottomrule
\end{tabular}
\end{table}

\begin{table}[!ht]
\centering
\caption{Comparison of total injected ($P_{\text{inj}}$) and stored ($P_{\text{stor}}$) power in (kW) under FLC/CPFL(50/50), DPFL/CPFL(50/50) and fully CPFL.}
\label{tab:power_results_transposed}
\scriptsize
\resizebox{\columnwidth}{!}{%
\begin{tabular}{c|c|c|c|c}
\toprule
\textbf{Fault Scenario} & \textbf{Power} & \textbf{FLC/CPFL} & \textbf{DPFL/CPFL} & \textbf{CPFL} \\
\hline
\multirow{2}{*}{F2} & $P_{\text{inj}}$  & 66.92 & 33.81 & 161.73 \\
                    & $P_{\text{stor}}$ & 103.44 & 56.19 & 347.64 \\
\hline
\multirow{2}{*}{F3} & $P_{\text{inj}}$  & 124.25 & 56.90 & 81.62 \\
                    & $P_{\text{stor}}$ & 325.28 & 228.10 & 317.94 \\
\hline
\multirow{2}{*}{F4} & $P_{\text{inj}}$  & 39.90 & 6.87 & 16.16 \\
                    & $P_{\text{stor}}$ & 250.92 & 204.30 & 240.27 \\
\hline
\multirow{2}{*}{F5} & $P_{\text{inj}}$  & 74.94 & 37.70 & 168.53 \\
                    & $P_{\text{stor}}$ & 106.95 & 54.77 & 355.08 \\
\bottomrule
\end{tabular}
}
\end{table}

The responses for FLC~50\% and DPFL~50\% under fault F2 are illustrated in Fig.~\ref{fig:flc_vs_dpfl_f2}. FLC~50\% quickly restores all controlled generators to the angular frequency margin $|\omega_i| \leq 0.01$, achieving stability significantly faster than DPFL~50\%, which exhibits oscillations and delayed synchronization. Additionally, FLC~50\% provides a smoother power control profile, avoiding the high-frequency fluctuations of DPFL~50\% which can accelerate battery degradation. 

\subsection{Comparative Energy Storage Utilization Across Control Schemes} \label{case_study_2}

In this study, we compare total power injection and storage across distributed control penetration 50\% for faults F2--F5, and the results are reported in Table~\ref{tab:power_results_transposed}. DPFL~50\% consistently yields the lowest values, reflecting its conservative strategy and minimal flexibility use. CPFL generally incurs the highest values, highlighting its aggressive, resource-intensive corrections. FLC~50\% balances between both, offering stronger responses than DPFL with lower energy demand than CPFL. However, in F3 and F4, FLC~50\% exceeds CPFL in energy usage, revealing inconsistent behavior and over-reliance on the ESS. These outcomes underscore the trade-offs: DPFL is conservative, CPFL aggressive, and FLC generally balanced.

\subsection{Limitations}

As demonstrated in Section~\ref{case_study}, while the proposed FLC framework exhibits generalization and competitive performance at moderate levels of distributed control penetration, its stability benefits diminish compared to DPFL at higher penetrations (beyond 70\%). This highlights the need for improved training and federated aggregation strategies to maintain system resilience under extensive distributed deployments.

The computational complexity of the proposed Cheby KAN model was evaluated using the \texttt{calflops} package, yielding $1,958$ floating-point operations per second (FLOPS) per input pass with a total of $768$ trainable parameters. While this represents a lightweight parameterization compared to conventional deep neural networks, the average simulation execution time of \( F_{1} \)–\( F_{5} \) for $t_{\text{max}}=300$ seconds indicates significant overhead: $184.51$ seconds for FLC~50\% compared to only $2.20$ seconds for DPFL~50\%. These results highlight that the inference speed of the Cheby KAN model remains a bottleneck. Enhancing and optimizing model execution is therefore critical to enable real-time deployment in power system control applications.

\section{Conclusions and Future Work}



This work presented the FLC multi-agent framework as a hybrid centralized--distributed control strategy for enhancing transient stability in power systems. It utilizes interpretable Chebyshev KAN-based neural controllers trained on centralized actions and deployed for distributed execution. The framework was evaluated on the IEEE 39-bus, 10-generator New England system across various penetration levels (10\%--80\%). Results show that FLC generalizes well to unseen faults, especially at moderate penetrations (10\%--50\%), consistently outperforming DPFL in both stability time and smoother power injection/absorption profiles. Performance degrades beyond 60\%, indicating challenges in maintaining distributed coordination at scale. Overall, FLC offers a promising balance of generalization, adaptability, and resilience for modern SGs.

Future research will tackle current reported limitations, and investigate FLC's performance across a broader set of contingencies and the integration of RL agents to enhance adaptivity and generalization. Additionally, given the relatively high inference time of Cheby KAN models, efforts will focus on optimizing its architecture and accelerating its execution to ensure practical feasibility for real-time control applications deployments.


\bibliographystyle{IEEEtran}
\bibliography{bibfile}

\end{document}